%% file: main.tex
\documentclass[10pt,letterpaper]{article}
\usepackage[top=0.85in,left=2.75in,footskip=0.75in]{geometry}

\usepackage{amsmath,amssymb}

\usepackage{changepage}

\usepackage{textcomp,marvosym}

\usepackage{cite}

\usepackage{nameref,hyperref}

\usepackage[right]{lineno}

\usepackage[nopatch=eqnum]{microtype}
\DisableLigatures[f]{encoding = *, family = * }

\usepackage[table]{xcolor}

\usepackage{array}

\newcolumntype{+}{!{\vrule width 2pt}}

\newlength\savedwidth

\usepackage{setspace} 
\raggedright
\usepackage[aboveskip=1pt,labelfont=bf,labelsep=period,justification=raggedright,singlelinecheck=off]{caption}

\makeatletter
\renewcommand{\@biblabel}[1]{\quad#1.}
\makeatother

\usepackage{lastpage,fancyhdr,graphicx}
\usepackage{epstopdf}
\fancyheadoffset[L]{2.25in}
\fancyfootoffset[L]{2.25in}

\usepackage{booktabs}
\usepackage{multirow}
\usepackage{vcell}
\usepackage{CJKutf8}
\newcommand{\chinese}[1]{\begin{CJK}{UTF8}{gbsn}#1\end{CJK}}
\usepackage{tabularray}
\usepackage{soul}

\begin{document}
\vspace*{0.2in}

\begin{flushleft}
{\Large
\textbf\newline{Warning: humans cannot reliably detect speech deepfakes} 
}
\newline
\\
Kimberly T. Mai\textsuperscript{1,2,*},
Sergi Bray\textsuperscript{1,2},
Toby Davies\textsuperscript{1},
Lewis D. Griffin\textsuperscript{2}
\\
\bigskip
\textbf{1} Department of Security and Crime Science, University College London, London, United Kingdom
\\
\textbf{2} Department of Computer Science, University College London, London, United Kingdom
\\
\bigskip

%
%





* Corresponding author\\
E-mail: kimberly.mai@ucl.ac.uk (KTM)

\end{flushleft}
\section*{Abstract}
Speech deepfakes are artificial voices generated by machine learning models. Previous literature has highlighted deepfakes as one of the biggest security threats arising from progress in artificial intelligence due to their potential for misuse. However, studies investigating human detection capabilities are limited. We presented genuine and deepfake audio to n = 529 individuals and asked them to identify the deepfakes. We ran our experiments in English and Mandarin to understand if language affects detection performance and decision-making rationale. We found that detection capability is unreliable. Listeners only correctly spotted the deepfakes 73\% of the time, and there was no difference in detectability between the two languages. Increasing listener awareness by providing examples of speech deepfakes only improves results slightly. As speech synthesis algorithms improve and become more realistic, we can expect the detection task to become harder. The difficulty of detecting speech deepfakes confirms their potential for misuse and signals that defenses against this threat are needed.

\input{1_introduction.tex}
\input{2_related_work.tex}
\input{3_method.tex}
\input{4_results.tex}
\input{5_limitations.tex}
\input{6_discussion.tex}
\input{7_supporting.tex}

\section*{Acknowledgments}
Kimberly T. Mai thanks Kelvin Ma for assistance with translating the survey materials.

\bibliography{bibliography}
\end{document}

%% file: 1_introduction.tex
\section*{Introduction}
Adversaries are already using speech deepfakes to commit fraud. In 2020, a bank manager in Hong Kong received a phone call from someone sounding like a company director he had spoken to before \cite{brewster_2022}. The purported director requested the bank manager to authorize transfers totaling \$35 million.  Based on their existing relationship, the bank manager transferred \$400,000 until he realized something was wrong.   The bank manager was a victim of an elaborate hoax: fraudsters had used deepfake technology to clone the director's voice. This incident is not isolated. In 2019, the CEO of a UK-based firm was swindled by a speech deepfake of his manager into transferring €220,000 to a Hungarian supplier \cite{stupp_2019}. 

Speech deepfakes are artificial voices generated by machine learning models. Due to rapid research progress, it is possible to produce a realistic-sounding clone using only a few audio samples \cite{choi2020}. This development raises the prospect of exploiting speech deepfakes for various criminal activities. Alongside impersonation, criminals may use deepfakes for spear phishing, propagating fake news, and bypassing biometric authentication systems \cite{alspach_2022, caldwell2020ai, mirsky2022threat}. 

Existing speech deepfake detection research focuses on developing machine learning systems in the context of voice authentication \cite{wu2017, nautsch2019, yamagishi2021asvspoof}. Comparisons beyond this biometric setting and studies which measure human detection capabilities are sparse \cite{gamage2022deepfakes}. 

The state of existing research raises questions. Firstly, machine learning systems require large amounts of data for training \cite{Goodfellow-et-al-2016} and are hard to interpret \cite{zhang2021survey}. When analyzing these systems, it is unclear which characteristics distinguish synthesized speech from bona fide. Therefore, knowing what humans use to identify deepfakes could provide a better understanding of how black-box machine learning systems work.

Secondly, focusing on automated biometric authentication does not quantify the threat of other potential criminal applications of speech deepfakes. Understanding the extent of these threats is critical. Multiple studies deem other uses of speech deepfakes as more concerning, such as misleading people through voice impersonations \cite{caldwell2020ai, mirsky2022threat}. Experts expect disinformation from deepfakes to erode trust on several levels: towards individuals, organizations, and even societies \cite{van2021tackling}. Moreover, it is estimated as much as 90\% of online content will be synthetically generated by 2026 \cite{schick2020deep}, meaning it will be challenging to moderate what gets produced. Therefore, understanding the risks of speech deepfakes will enable the development of better defenses and regulations to counteract hazards before they occur.

We seek to address these two questions by measuring how well humans distinguish bona fide speech from synthesized speech. We ran an online experiment where individuals listened to bona fide and fake audio clips and attempted to differentiate between them.

We randomly assigned the participants to two configurations. In the first configuration, we presented participants with one audio clip at a time and asked them to decide if the clip was fake. In the second configuration, we presented participants with audio clip pairs containing the same speech (one bona fide and one synthesized) and asked them to identify the synthesized audio. 

We ran the experiment in English and Mandarin to understand if listeners used language-specific attributes to detect deepfakes and to observe if deepfake detection is more manageable in one language than another. Finally, we incorporated randomized interventions to evaluate whether familiarizing participants with examples of speech deepfakes boosts detection performance.

Our results suggest the listeners had limited detection capabilities, and performance is similar between languages. Additionally, familiarizing participants improved performance but only to a small extent.

%% file: 2_related_work.tex
\section*{Background}

\subsection*{Deepfake media}

Deepfakes are synthetic media produced in the likeness of a person. They fall under the field of generative artificial intelligence (AI). Generative AI is a subset of machine learning (ML) algorithms that learn the patterns and characteristics of a dataset \cite{Goodfellow-et-al-2016}. The algorithms use this knowledge to generate synthetic content similar to the original data. Deepfakes specifically refer to the outputs of generative AI that resemble humans and their actions.

Deepfake media occur in different modalities:

\begin{enumerate}
    \item \textbf{Images}: This modality contains static faces generated using varying techniques. These techniques include:
        \begin{itemize}
            \item \textbf{Generation from scratch}: A generative adversarial network \cite{goodfellow_gan} or diffusion model \cite{sohl2015} synthesizes a fictional identity.
            \item \textbf{Morphing}: Blending similar-looking faces to produce an identity containing the characteristics of the sources \cite{damer2018}.
            \item \textbf{Swaps}: A source face replaces the target in a different image \cite{bitouk2008}.
        \end{itemize}
    \item \textbf{Video}: This modality features individuals performing actions. Currently, the techniques used to synthesize videos are similar to those used in images. Image synthesis techniques are applied at a frame level and stitched together to form a video.
    \item \textbf{Speech}: This modality conveys information in a manner that sounds like a genuine person's voice. Although audio can refer to general sound synthesis, the terms ``audio", ``speech", and ``voice" deepfakes are used interchangeably in academic literature. We refer to them as ``speech deepfakes" for consistency throughout the text.
\end{enumerate}

In addition, deepfakes are either produced in the likeness of a known identity (\textbf{targeted}) or do not resemble a familiar identity (\textbf{untargeted}). For example, we can categorize video deepfakes of politicians as targeted. Conversely, a generic face created from scratch and not conditioned to resemble a specific individual is untargeted.

We refer the reader to Zhang (2022) \cite{zhang2022deepfake} for further information on deepfake terminology. As fewer works focus on speech deepfakes, we concentrate on this modality.

\subsection*{Synthesizing speech}
Generative models are often used to synthesize speech. Speech synthesizers which use generative models follow a common framework:

\begin{enumerate}
\item \textbf{Data collection}: Several audio recordings of the speaker are collected.
\item \textbf{Pre-processing}: The audio recordings are converted into alternative formats to make it easier for the generative model to work with them.
\item \textbf{Training}: Processed audio recordings are fed to the generative model to learn the patterns and characteristics of the data. The trained model is often called a vocoder.
\end{enumerate}

The frameworks often include text-to-speech (TTS) modules to make it easier to generate speech. The generative model also sees text transcriptions corresponding to the audio recordings in this setting.

\begin{figure}[!h]
\includegraphics[width=\linewidth]{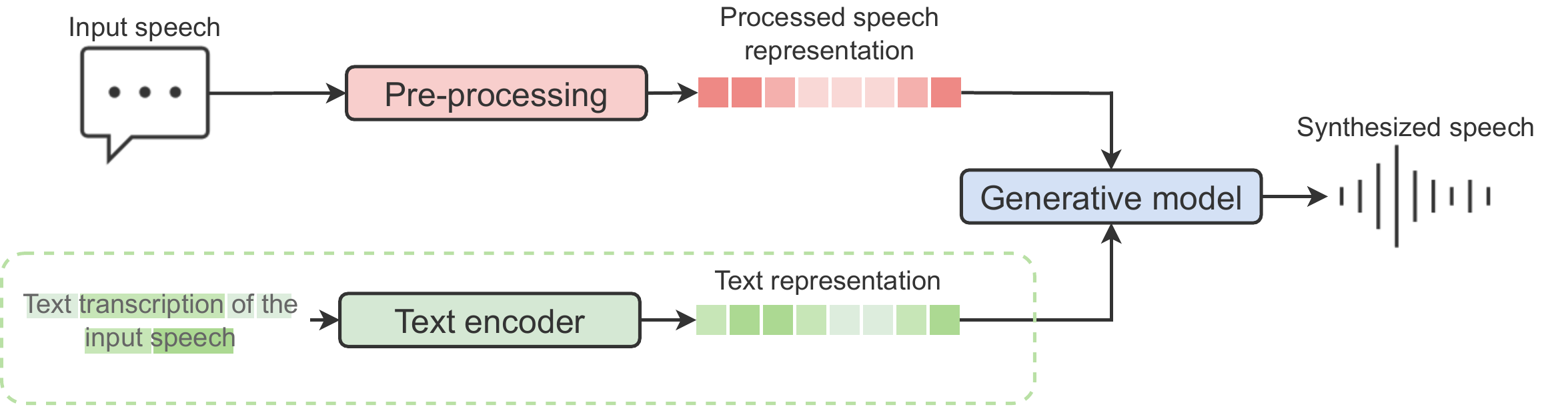}
\caption{{\bf Diagram of a typical generative speech synthesis model.}}
\label{fig1}
\end{figure}

We depict a visualization of this framework in Fig \ref{fig1}.

\subsection*{Related work on human deepfake detection capabilities}

Most deepfake detection studies which examine human performance use visual media. When faced with deepfake content of politicians, participants rely on contextual knowledge in the form of political literacy to identify spoofs \cite{barari_lucas_munger_2021, appel_2022}. 

Removing such background knowledge makes the detection task more difficult. In the context of images, multiple studies show humans do not perform much better than chance \cite{bray2022testing, nightingale2022}. There is no improvement when evaluating videos either \cite{groh2022, kobis2021, tahir2021seeing}. Moreover, these studies suggest humans are overconfident in their deepfake detection abilities \cite{kobis2021}.

Several of the above studies examine if interventions can boost detection performance. However, the effectiveness of these interventions is debatable. Bray et al. \cite{bray2022testing} familiarized participants by showing examples of deepfakes before the main task. The authors also drew participants' attention to errors often present in bogus images. Although these interventions improved deepfake detection performance, they also increased overall skepticism as a higher proportion of bona fide images were falsely classified. One could also note that pointing out errors biases the participants and prevents them from independently identifying the tell-tale characteristics of deepfakes. Köbis et al. \cite{kobis2021} presented interventions by informing participants about the impact of deepfakes and rewarding correct guesses. Neither intervention led to improved performance.

In contrast, other authors found interventions derived from ML model outputs improve detection. Tahir et al. \cite{tahir2021seeing} produced educational material containing indicators of bogus images with the assistance of ML interpretability tools. The authors found detection performance improved compared to the initial control group. However, a recent study \cite{geirhos2023viz} contests the reliability of these tools, as the authors show it is possible to manipulate the output visualizations. Groh et al. \cite{groh2022} allowed participants to amend their choices after viewing the predictions of an ML model. This form of cooperation improved results significantly.

There are fewer studies that examine how well humans can detect speech deepfakes. Watson et al. \cite{watson2021audio} presented eight clips to college students and asked them to decide whether the clips were real or fake. They found that shorter clips were easier to identify. However, the sample size of their study was small and skewed towards a younger, college-educated demographic. The ASVspoof challenge organizers ran an experiment with a larger sample size \cite{wang2020asvspoof}. They asked 1,145 participants to imagine they worked in a call center and decide whether the incoming calls were spoken by humans or by an AI. However, the experiment is limited to the speaker verification setting. 

M{\"u}ller et al. \cite{muller2022human} ran a game where 378 participants competed against an ML model to decide if an audio clip was fake. Similarly to Groh et al. \cite{groh2022}, they found that feedback from the ML model improved human performance. In their experiment, M{\"u}ller et al. \cite{muller2022human} found that the difference between human and AI accuracy was about 10\%. However, their study only used English-language clips, only presented one audio clip to participants at a time, and did not collect information about participant confidence.

We summarize the relevant literature in Table \ref{tab:literaturereview}. We note that Barari et al. \cite{barari_lucas_munger_2021} mention fake speech stimuli in their analysis. However, they used actors to create the speech instead of generative AI. Therefore we excluded this from our analysis.

\begin{table}[h!]
\begin{adjustwidth}{-2.5 in}{0in}
\centering
\caption{\bf{Summary of related literature measuring human capabilities to detect deepfakes.}}
\label{tab:literaturereview}
\begin{tblr}{
  cell{2}{1} = {r=2}{},
  cell{4}{1} = {r=5}{},
  cell{9}{1} = {r=3}{},
  hline{1,12} = {-}{0.08em},
  hline{2,4,9} = {-}{0.05em},
}
Modality & Year & Author             & Deepfake stimuli                                                                                          \\
Image    & 2021 & Nightingale \& Farid \cite{nightingale2022} & Faces generated using StyleGAN2 \cite{karras2020analyzing}                                                                            \\
         & 2022 & Bray et al. \cite{bray2022testing}        & Faces generated using StyleGAN2                                                                            \\
Video    & 2021 & Barari et al. \cite{barari_lucas_munger_2021}      & Face-swap videos of politicians                                                                           \\
         & 2021 & Groh et al. \cite{groh2022}       & {Face-swap videos from the Deepfake Detection Challenge dataset \cite{dolhansky2020deepfake}}                                        \\
         & 2021 & Köbis et al. \cite{kobis2021}      & {Face-swap videos from the Deepfake Detection Challenge dataset}                                        \\
         & 2021 & Tahir et al. \cite{tahir2021seeing}      & {Face-swap videos from Celeb-DF \cite{li2020celeb}, FaceForensics++ \cite{rossler2019faceforensics++} and DeepFaceLab \cite{perov2020deepfacelab}}                                       \\
         & 2022 & Appel \& Prietzel \cite{appel_2022}   & Face-swap videos of politicians                                                                           \\
Speech   & 2020 & Wang et al. \cite{wang2020asvspoof}       & {\footnotesize{Spoofed utterances generated from TTS and voice conversion systems used in ASVspoof2019}} \\
         & 2021 & Watson et al. \cite{watson2021audio}      & Audio clips generated using MelGAN \cite{kumar2019melgan}                                                                        \\
         & 2022 & M{\"u}ller et al. \cite{muller2022human}     & {Spoofed utterances from the ASVspoof2019 dataset \cite{wang2020asvspoof}}                                                      
\end{tblr}
\end{adjustwidth}
\end{table}

\clearpage

%% file: 3_method.tex
\section*{Materials and methods}

Our research questions were as follows:

\begin{enumerate}
    \item How well can humans detect speech deepfakes?
    \item Are there differences in detection capabilities depending on the language?
    \item Do interventions in the form of examples and added context improve detection performance?
\end{enumerate}

Our experiments focused on human performance rather than the performance of automated detectors. Through this setup, we could quantify the threat of speech deepfakes when humans interact with them.

\subsection*{Stimuli}

\subsubsection*{Bona fide stimuli}
We collected bona fide stimuli from two publicly available datasets. Both datasets consist of one female speaker reading generic sentences. The datasets also include text transcriptions of the audio. We chose such datasets to prevent participants from using external cues for the detection task.

We used LJSpeech \cite{ljspeech17} as the English dataset. The dataset consists of a speaker reading passages from seven non-fiction books, varying between one and ten seconds in length.

We used the Chinese Standard Mandarin Speech Corpus (CSMSC) \cite{csmsc2019} as the Mandarin dataset. The corpus used in the dataset aims to cover Mandarin tones and prosody as comprehensively as possible.

\subsubsection*{Deepfake stimuli}
To create the deepfake stimuli, we used publicly available TTS models trained on the two datasets \cite{watanabe2018espnet}. In particular, we chose pre-trained VITS models \cite{kim2021conditional}. VITS is an end-to-end TTS model which combines the data pre-processing and vocoder into a single framework.

We randomly selected 50 sentences from the validation split of the two datasets to create the deepfakes. We used the same sentences for our bona fide stimuli. Therefore, we had 100 clips in total.

The validation split consists of samples not used for training the ML models. It is good practice to use unseen data because it indicates how well a trained model generalizes. Consequently, the resultant generated audio should contain artifacts that we would expect to hear from ML models. These artifacts might serve as informative features for distinguishing deepfakes. If we used samples previously seen during training, the model could potentially mimic the samples perfectly and would not contain representative artifacts.

\subsection*{Procedure}
\begin{figure}[!h]
\includegraphics[width=\linewidth]{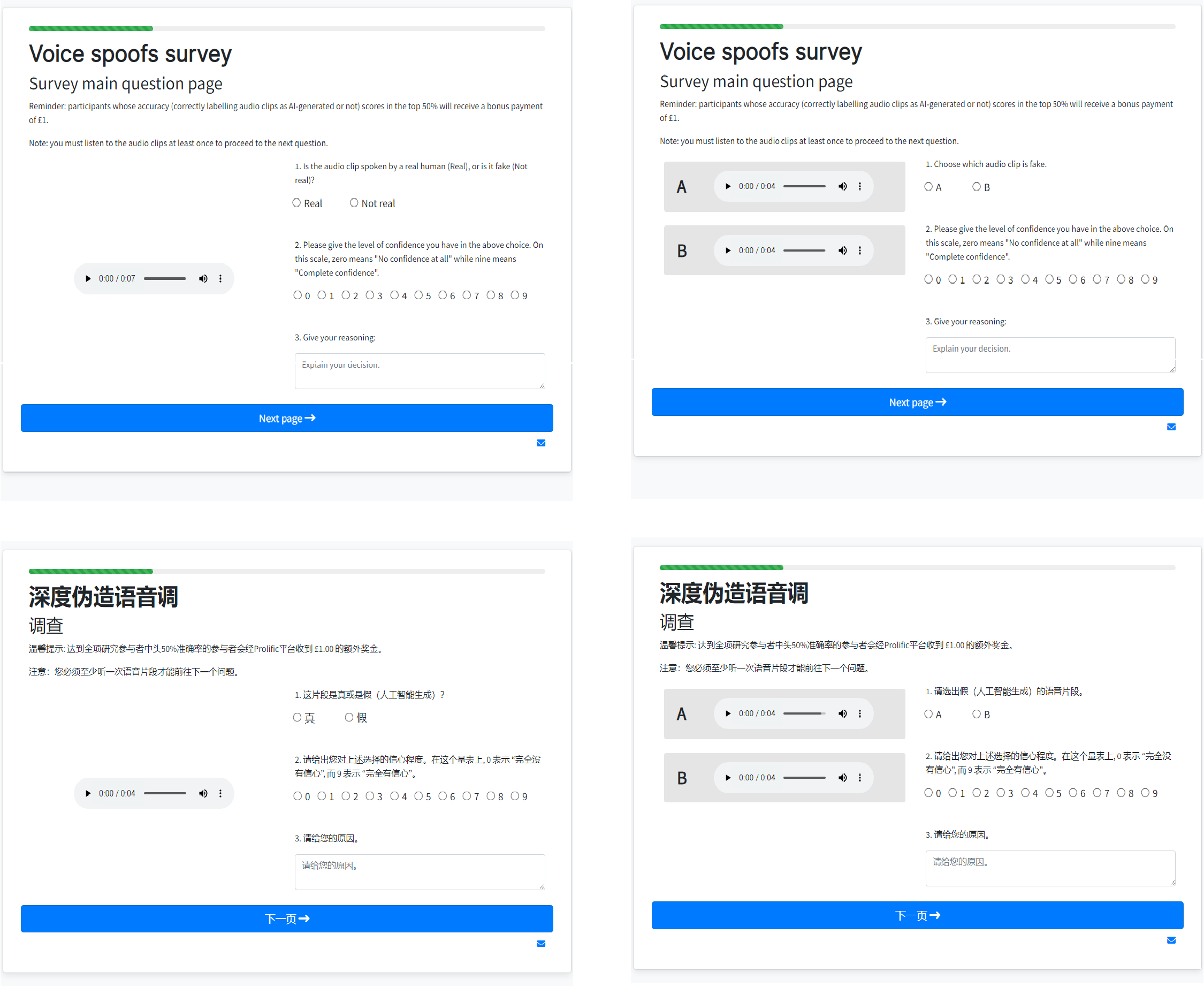}
\caption{{\bf Screenshots of the task interface.}}
\label{fig2}
\end{figure}

The setup for the English and Mandarin experiments was identical. We randomly assigned participants to two configurations: unary and binary. In both configurations, we asked participants to rate the confidence of their choice on a ten-point Likert scale and provide freeform text justifications. Participants were allowed to listen to the clips as often as they liked. We did not give feedback to the participants to inform them if their choices were correct. Compared to the setups described in M{\"u}ller et al. and Groh et al. \cite{muller2022human, groh2022}, the lack of feedback creates a more realistic scenario. When encountering speech deepfakes in the wild (for example, through fraudulent calls), humans do not know that the voices are fake. We include screenshots of the two configurations in Fig \ref{fig2}.

\newpage

\subsubsection*{Unary}
We presented 20 randomly chosen distinct clips to each participant, each on separate pages. Participants listened to approximately an equal number of bona fide and synthesized clips, but we did not inform them about the proportion. We tasked the participants with deciding whether the clip they heard was real or fake.

\subsubsection*{Binary}
We presented 20 randomly chosen clip pairs (labeled `A' and `B') comprising the same spoken sentence. Each pair contained a clip uttered by the human speaker and a clip produced by VITS. We randomized the order of the fake and real clips and asked the participants to decide which clip was fake. We included this scenario to see if contextual information helped detection.

\subsubsection*{Familiarization treatment}
In addition to the two configurations, we randomly assigned half of the participants to a familiarization treatment group. We included the treatment to verify the existing literature and understand if humans could be trained to detect deepfakes like an ML model. We showed participants in the treatment group five deepfake utterances before commencing the main detection task. We informed the participants that these examples were synthesized and allowed them to listen to the clips multiple times. These clips were distinct from the stimuli used in the main task.

For participants in the control group, we gave them a filler task. In this task, we asked participants to list potential applications of synthesized speech and to provide their opinion about whether synthesized audio will positively or negatively impact society.

\subsection*{Participants}
We recruited participants via the Prolific platform. We filtered for participants fluent in English and Mandarin, as fluency affects detection performance \cite{muller2022human}. We paid participants at a rate of £7.25 per hour. To encourage more thoughtful responses, we informed participants they could receive a £1.00 bonus if their detection scores were in the top 50\%. Overall, we recruited 529 participants. The mean age was 28.9 years old, and 50.6\% identified as male. Table \ref{table1} contains a more detailed breakdown of the demographics by treatment group.

\begin{table}[!ht]
\centering
\caption{\bf {Number of participants by group.}}
\label{table1}
\begin{tabular}{lllllll} 
\toprule
Group                     & \multicolumn{3}{l}{English}                         & \multicolumn{3}{l}{Mandarin}                         \\ 
\cline{2-7}
                          & \textit{n} & \textit{Age (SD)} & \textit{Male (\%)} & \textit{n} & \textit{Age (SD)} & \textit{Male (\%)}  \\ 
\hline
Unary no familiarization  & 76         & 26.8 (8.1)      & 55.2               & 65         & 31.0 (10.4)     & 44.6               \\
Unary familiarization     & 65         & 26.7 (7.3)      & 56.9               & 54         & 31.4 (8.7)      & 44.4                \\
Binary no familiarization & 60         & 27.5 (7.2)      & 53.3               & 70         & 31.8 (9.0)      & 48.8                \\
Binary familiarization    & 80         & 27.4 (7.3)      & 57.5               & 59         & 29.1 (8.5)      & 39.6                \\ 
\hline
Overall                   & 281        & 27.1 (7.5)      & 55.8               & 248        & 30.9 (9.2)      & 44.5                \\
\bottomrule
\end{tabular}
\end{table}

\subsection*{Ethics statement}
The study was reviewed and exempted by the Department of Security and Crime Science's ethics board at University College London. All participants were notified about the purpose of the study and were over the age of 18. Prior to participating, the participants were asked to tick a series of checkboxes to provide informed written consent.

\subsection*{Benchmarking against automated deepfake detectors}

To compare the performance of the human participants to automated methods, we trained two artificial neural networks which specialized in detecting speech deepfakes. The two networks used an LFCC-LCNN architecture \cite{wang2021interspeech}. LFCC-LCNNs convert raw audio waveforms into two-dimensional representations. They learn by seeing bona fide and deepfake samples and are rewarded for correctly classifying a sample's authenticity. The ASVspoof 2021 challenge used LFCC-LCNNs as baseline models for spoof detection \cite{yamagishi2021asvspoof}. Hence, they are a reasonable benchmark for our experiments. For more detail about the top-performing speech deepfake detection architectures, we refer the reader to the article summarizing ASVspoof 2021 \cite{yamagishi2021asvspoof}.

We used two versions for each language:
\begin{enumerate}
    \item In-domain: We trained the networks using the training split of LJSpeech and CSMSC as bona fide samples and created deepfakes by passing the sentences of the training splits through VITS.
    \item Out-of-domain: We trained the Mandarin network with FAD \cite{ma2022fad}, another Mandarin-language dataset. We used the pre-trained ASVspoof network \cite{delgado2021asvspoof} for English-language evaluation.
\end{enumerate}

Artificial neural networks are known to perform well when evaluating against samples similar to those seen during training. However, their performance often drops when encountering different examples \cite{shen2021towards}, even if they are in the same language. These differences can be subtle to a human listener and include changes in the speaker's identity or environment. Therefore, we introduce the out-of-domain version for a fairer comparison with human performance, especially as it is unlikely that the participants in our study recognize the LJSpeech and CSMSC identities.

%% file: 4_results.tex
\section*{Results}

\subsection*{Overall performance}

\begin{figure}[!h]
\includegraphics[width=\linewidth]{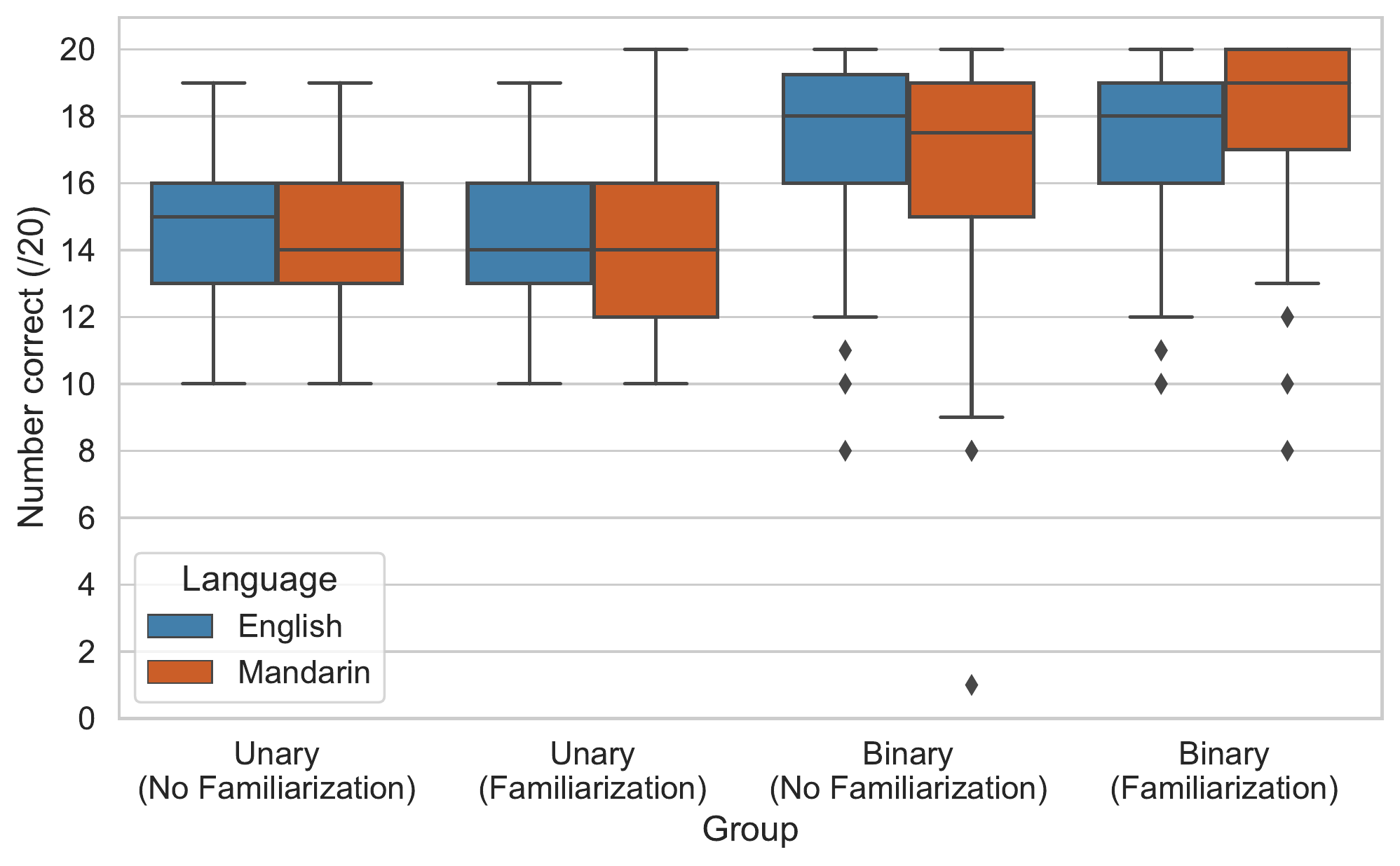}
\caption{{\bf Box plot summarizing human performance across the different groups.}}
\label{fig3}
\end{figure}

Fig \ref{fig3} summarizes human performance across all of the different groups. We provide breakdowns of the classification choices in Tables \ref{table2} and \ref{table3}, which aggregate the English and Mandarin results. We completed the analysis using the SciPy \cite{2020SciPy-NMeth} and statsmodels \cite{seabold2010statsmodels} Python packages. For further details, the \nameref{supporting} contains results per stimulus. 

\begin{table}[!ht]
\caption{\bf{Confusion matrix for the unary group responses.}}
\label{table2}
\begin{tabular}{llll} 
\cmidrule[\heavyrulewidth]{3-4}
\multicolumn{2}{l}{\multirow{2}{*}{}}              & \multicolumn{2}{l}{Predicted class}          \\ 
\cmidrule{3-4}
\multicolumn{2}{l}{}                               & \textit{Real (2,442)} & \textit{Fake (2,678)}  \\ 
\midrule
\multirow{2}{*}{True class} & \textit{Real (2,598)} & 1,761                 & 837                   \\
                            & \textit{Fake (2,522)} & 681                  & 1,841                  \\
\bottomrule
\end{tabular}
\begin{flushleft} \textit{n} = 5,120.\\
            Overall accuracy = 70.35\%.\\
            Reals correctly identified = 67.78\%.\\
            Fakes correctly identified = 73.0\%.
\end{flushleft}
\end{table}

\begin{table}[!h]
\caption{\bf{Confusion matrix for the binary group responses.}}
\label{table3}
\begin{tabular}{llll} 
\cmidrule[\heavyrulewidth]{3-4}
\multicolumn{2}{l}{\multirow{2}{*}{}}              & \multicolumn{2}{l}{Predicted class}  \\ 
\cmidrule{3-4}
\multicolumn{2}{l}{}                               & \textit{Real} & \textit{Fake}        \\ 
\midrule
\multirow{2}{*}{True class} & \textit{Real}        & -             & -                    \\
                            & \textit{Fake (5,380)} & 775           & 4,605                 \\
\bottomrule
\end{tabular}
\begin{flushleft}
    True \textit{real} class labels are not defined in this scenario as participants were asked to choose the fake clip every time.\\
            Overall accuracy is equivalent to fakes correctly identified = 85.59\%.
\end{flushleft}
\end{table}

Participants made the correct classifications 70.35\% of the time in the unary scenario. They were better at identifying deepfakes (73\% accuracy). In comparison, participants correctly identified bona fide examples 67.78\% of the time. We speculate the high number of misclassified bona fide samples is partly due to increased skepticism, as participants were aware of the presence of deepfakes through the task briefing. This behavior aligns with observations in Bray et al. \cite{bray2022testing}.

Performance improved under the binary scenario. Participants correctly recognized the deepfake audio in 85.59\% of trials. However, the binary setup represents an unrealistic scenario. Even if the speaker's identity is known, reference utterances containing the same speech as the test clip we would like to evaluate are unlikely to be available.

\subsection*{Measuring the effects of interventions}

We follow a similar approach to Groh et al. \cite{groh2022} to disentangle the effects of each intervention on performance. We transformed the correct/incorrect results into continuous values by weighting each participant's decision with their provided confidence scores. 

The ten-point confidence scale participants completed serves as the mapping function. The lowest score of 0 signals that the participant's choice is a guess, so their confidence in making the right decision corresponds to 50\%. In contrast, the highest score of 9 corresponds to 100\% belief. 

The resulting transformed scores depended on whether the participants made the correct classification. For example, if the participant rated their confidence as 7, this maps to a belief of 88\%. If they make the right decision, the adjusted score is 0.88. Conversely, if they make the wrong decision, we subtract the value from 1, resulting in an adjusted score of 0.12.

The revised scores also enable fairer comparisons with the automated deepfake detectors, which output scores between 0 and 1 when evaluating examples. We refer to the revised scores as accuracy scores for the remainder of the text. We also rescale the scores to percentages.

After transforming the results, we analyzed the effects of different interventions on the accuracy scores of participants on each audio clip using linear regression. In addition to language, familiarization and binary intervention, we analyzed the impact of the clip duration. Table \ref{table4} outlines the results at the overall, unary and binary levels.

\begin{table}[!h]
\centering
\caption{\textbf{Linear regression results of interventions on confidence-scaled accuracy.}}
\label{table4}
\begin{tblr}{
  row{7} = {t},
  cell{1}{2} = {c=3}{},
  cell{3}{1} = {t},
  cell{4}{1} = {t},
  cell{5}{1} = {t},
  cell{6}{1} = {t},
  hline{1,12} = {-}{0.08em},
  hline{2} = {2-4}{0.03em},
  hline{3,8} = {-}{0.05em},
}
Independent variable        & Dependent variable: Confidence-scaled accuracy &                      &                      \\
                            & \textit{All (SD)}                              & \textit{Unary (SD)}  & \textit{Binary (SD)} \\
Constant                    & {43.742*** (1.897)}                           & {46.394*** (2.791)} & {71.217*** (2.530)} \\
Mandarin\textsuperscript{a} & {1.790 (1.404)}                               & {1.477 (1.882)}     & {2.152 (2.118)}     \\
Familiarization             & {3.840*** (1.191)}                            & {3.758** (1.571)}   & {3.854** (1.802)}   \\
Clip length                 & {0.797*** (0.209)}                            & {0.375 (0.358)}     & {1.168*** (0.230)}  \\
Binary intervention         & {29.830*** (1.186)}                           & -                    & -                    \\
Observations                & 10,500                                         & 5,120                & 5,380                \\
$R^2$                       & 0.165                                          & 0.003                & 0.010                \\
Adjusted $R^2$              & 0.165                                          & 0.002                & 0.009                \\
F-Statistic                 & 171.102***                                     & 2.455*               & 11.794***            
\end{tblr}
\begin{flushleft}
            \textsuperscript{a}Dummy variable indicating which language was used in the task. 1 = Mandarin, 0 = English.\\
            \textsuperscript{*}$p < 0.1$.
            \textsuperscript{**}$p < 0.05$.
            \textsuperscript{***}$p < 0.01$.
\end{flushleft}
\end{table}

\clearpage
\subsubsection*{Reference audio helps with deepfake detection}

The linear regression results indicate the improvement gained from the binary scenario is statistically significant ($p < 0.001$). Consequently, the results suggest contextual information via reference audio is beneficial for uncovering quirks in synthesized speech.

\subsubsection*{Training humans to detect deepfakes only helps slightly}

The familiarization treatment increases detection accuracy by 3.84\% on average ($p = 0.001$). This effect is also present in the unary and binary regression results, improving accuracy by 3.76\% ($p = 0.017$) and 3.85\% ($p = 0.032$), respectively. However, incorporating familiarizations equates only to an accuracy slightly above chance (52.31\%) in the unary setting for the mean clip length (5.76 seconds), ceteris paribus.

\subsubsection*{It is equally challenging to detect deepfakes in Mandarin and English}

Fig \ref{fig3} shows that performance in English and Mandarin is comparable across the different treatment groups. This observation is supported by Table \ref{table4}, which shows Mandarin-speaking participants only outperform their English counterparts by 1.79\%, and this effect is not statistically significant ($p = 0.202$).

\subsubsection*{Shorter speech deepfakes are not easier to identify}

\begin{figure}[h!]
\includegraphics[width=0.7\linewidth]{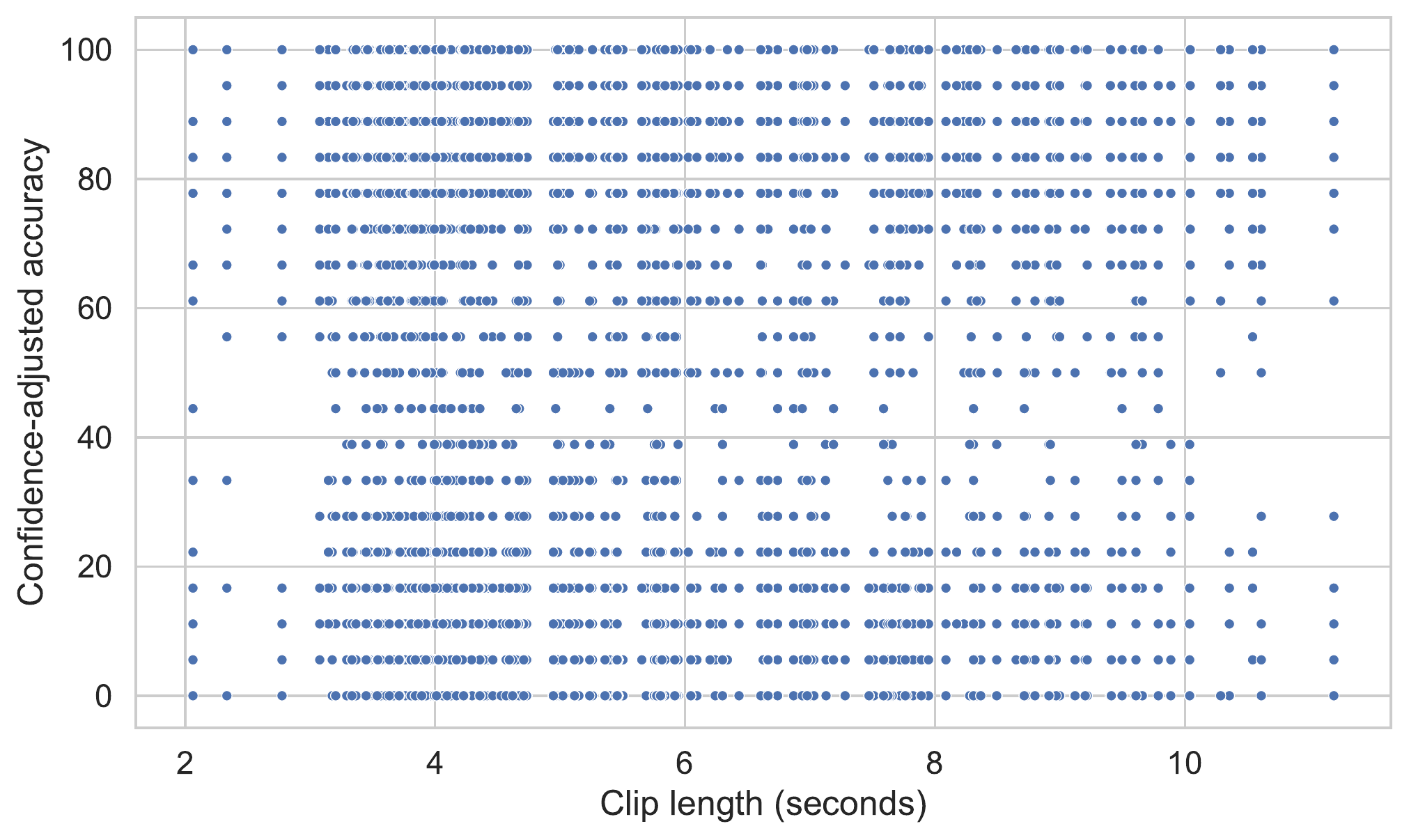}
\caption{{\bf Scatter plot showing the relationship between clip length and confidence-scaled accuracy.}}
\label{fig4}
\end{figure}

As our stimuli varied from 2 to 11 seconds, we included clip length in the regression to verify whether it is easier to discriminate shorter clips. Our results suggest clip length has a negligible impact on accuracy, improving performance by only 0.80\% for each additional second. Our scatter plot (Fig \ref{fig4}) supports this and shows no relationship between the two variables. These findings conflict with Watson et al. \cite{watson2021audio}, who suggest it is easier to identify shorter deepfakes.

\newpage
\subsection*{Analyzing performance against time}

In addition to analyzing the treatment effects, we examine whether the hypothesis of spending more time on the task improves performance.

\subsubsection*{Listening to the clips more frequently does not aid detection}

\begin{figure}[!h]
\includegraphics[width=0.7\linewidth]{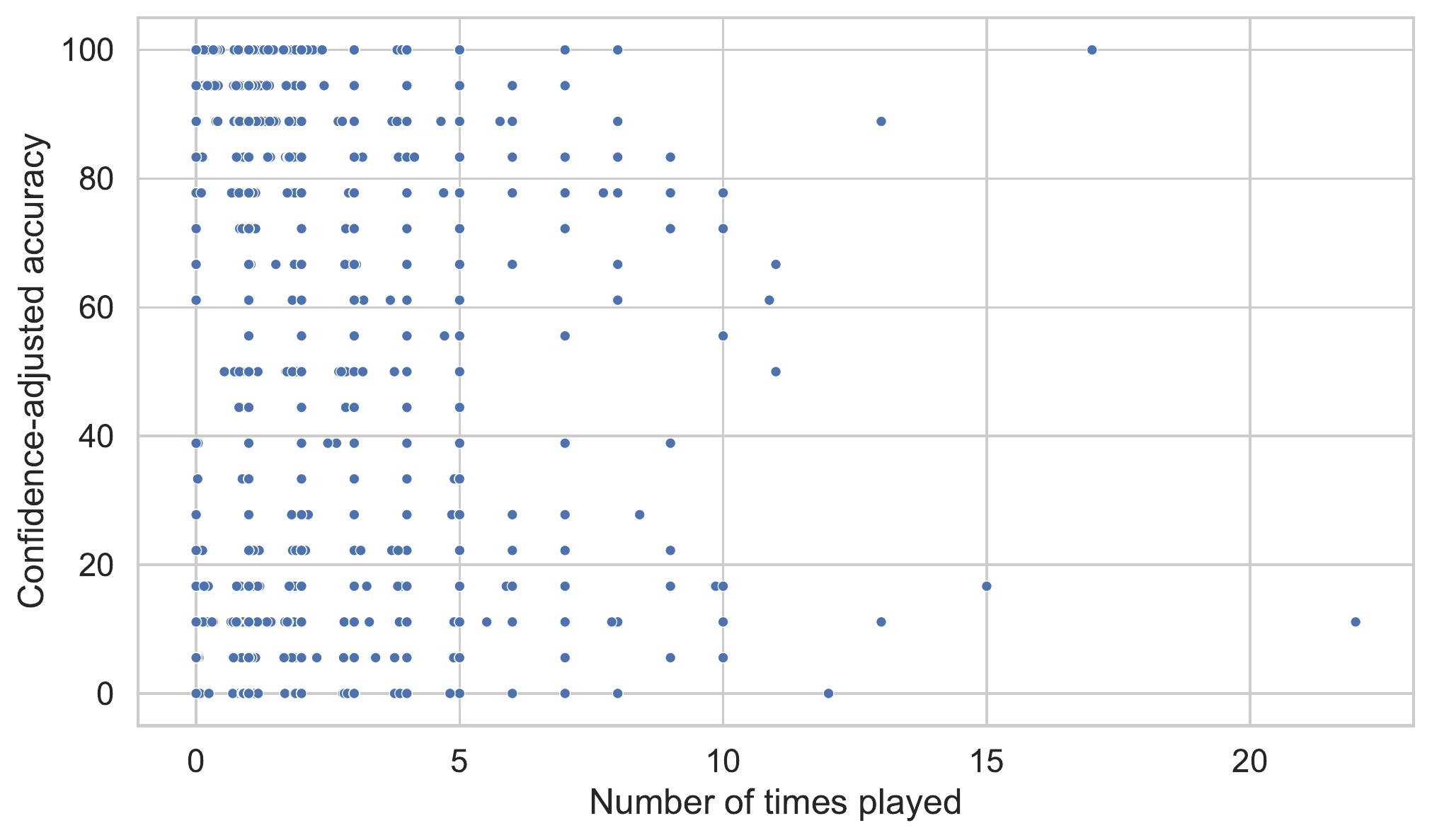}
\caption{{\bf Scatter plot showing the relationship between the number of times played and confidence-scaled accuracy.}}
\label{fig5}
\end{figure}

We recorded the number of times participants clicked on each audio clip and compared the values to accuracy. As shown in Fig \ref{fig5}, there is no relationship between the two variables ($\rho = -0.05$, $p < 0.001$).

\subsubsection*{Spending more time on the task also does not affect performance}

\begin{figure}[!htb]
\includegraphics[width=0.7\linewidth]{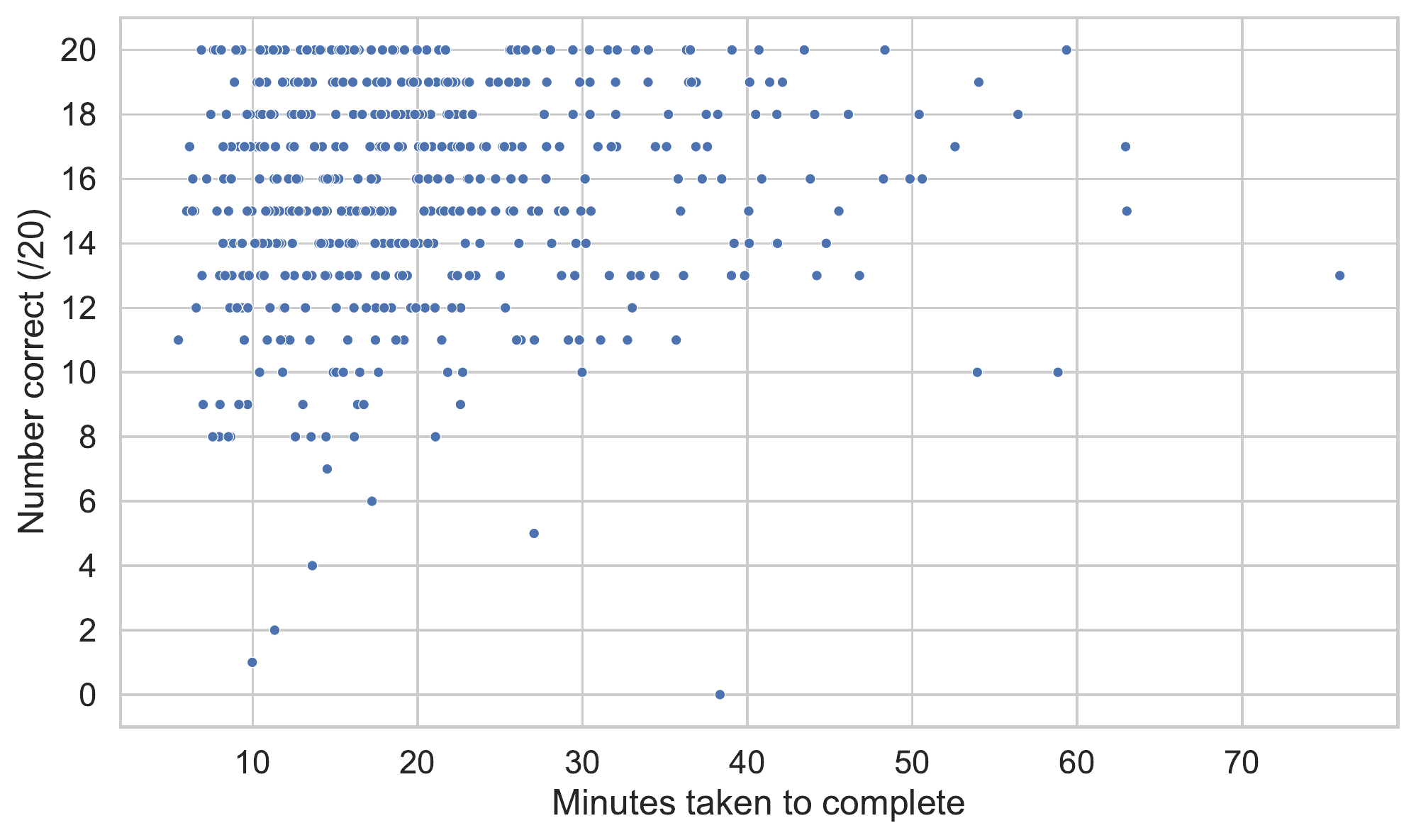}
\caption{{\bf Scatter plot showing the relationship between minutes taken to complete and correctness scores.}}
\label{fig6}
\end{figure}

Similar to the above analysis, we compared the time taken to complete the entire task to the total number of clips correctly identified. Fig \ref{fig6} does not indicate a relationship between the two variables ($\rho = 0.10$, $p = 0.018$), suggesting investing more time to complete the task does not improve performance.

\subsubsection*{Participants do not get better throughout the task without explicit feedback}

\begin{figure}[!h]
\includegraphics[width=0.7\linewidth]{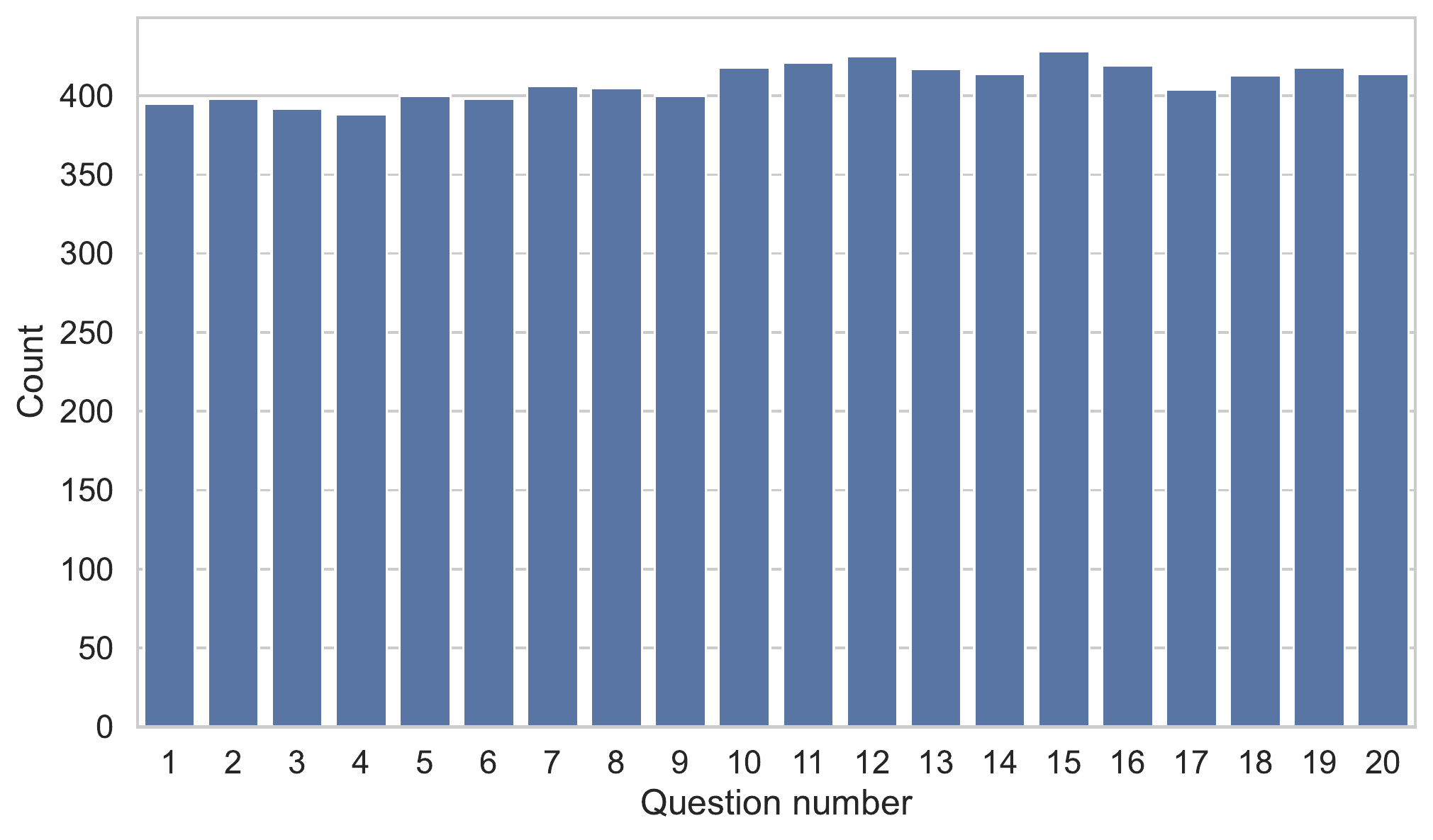}
\caption{{\bf Histogram of correct responses across question number.}}
\label{fig7}
\end{figure}

To understand whether participants improved as they saw more examples and progressed further in the task, we calculated the number of correct responses per question number. If so, we would expect more correct answers in question 20 compared to question 1. Fig \ref{fig7} illustrates the resulting histogram. The histogram shows performance is relatively stable across the questions. This observation indicates participants do not improve throughout the task unless they have explicit feedback, as examined by Groh et al. \cite{groh2022} and M{\"u}ller et al. \cite{muller2022human}. We quantitatively verified the result by conducting a one-way chi-squared hypothesis test against the uniform distribution, which was not statistically significant ($\chi^2 = 6.19$, $p = 0.997$).

\newpage

\subsection*{Comparing human performance to automated detectors}

The following section compares human performance to automated deepfake detectors. For comparability, we use commonly-reported performance metrics found in ML literature.

\begin{itemize}
    \item \textbf{Receiver operating characteristic (ROC)}: These plots represent discriminatory ability. They compare true positive rates against false positive rates at different thresholds.
    \item \textbf{The area under the receiver operating characteristic (AUROC)}: This score summarizes ROCs into a single value. 50\% AUROC indicates all predictions are guesses, whereas 100\% AUROC means perfect discrimination between bona fides and deepfakes in all trials.
    \item \textbf{Equal error rate (EER)}: This describes the point on ROCs where the true positive and false positive rates are equal.
\end{itemize}

\begin{figure}[!h]
\includegraphics[width=\linewidth]{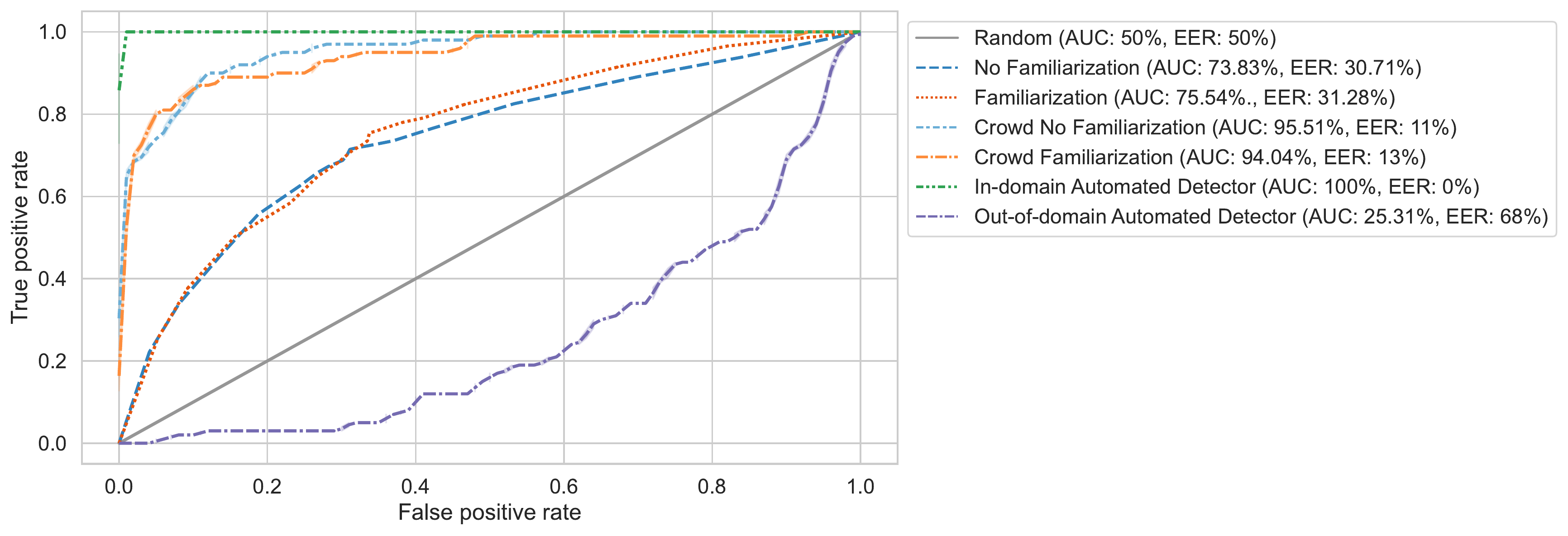}
\caption{{\bf Receiver operator curves under the unary scenario.}}
\label{fig8}
\end{figure}

Fig \ref{fig8} displays the AUROC and EER scores. We include only the unary scenario in this analysis as the inference setup between humans and automated detectors is more comparable. Both evaluate one clip at a time. We aggregated the English and Mandarin results as we observed similar results.

\subsubsection*{Human performance is less sensitive to unknown conditions compared to automated detectors}

The no familiarization (AUROC = 73.83\%) and familiarization curves (AUROC = 75.54\%) confirm humans performed better than chance. The curves also support the linear regression result. Showing participants examples of deepfakes only had a minute impact on performance.  However, performance was quite unreliable: on average, humans incorrectly classified clips a quarter of the time.

Humans underperformed the in-domain automated detectors, which had perfect discrimination ability (AUROC = 100\% for both languages). However, out-of-domain detectors often incorrectly classified bona fides as deepfakes (AUROC = 25.31\%). Based on this behavior, humans are more robust to unknown factors, such as speaker identity.

\subsubsection*{Crowd speech deepfake detection is comparable to the top-performing automated detectors}

Per Groh et al. \cite{groh2022}, we averaged participants' accuracy scores per clip to calculate the crowd-sourced responses. Like the results observed with video stimuli \cite{groh2022}, crowd performance is on par with the in-domain detector. However, the benefit of familiarizing participants dissipates when averaging responses. The crowd no familiarization and crowd familiarization AUROCs are similar at 95.51\% and 94.04\%, respectively.

\subsection*{Freeform text analysis}

\begin{figure}[!h]
\includegraphics[width=\linewidth]{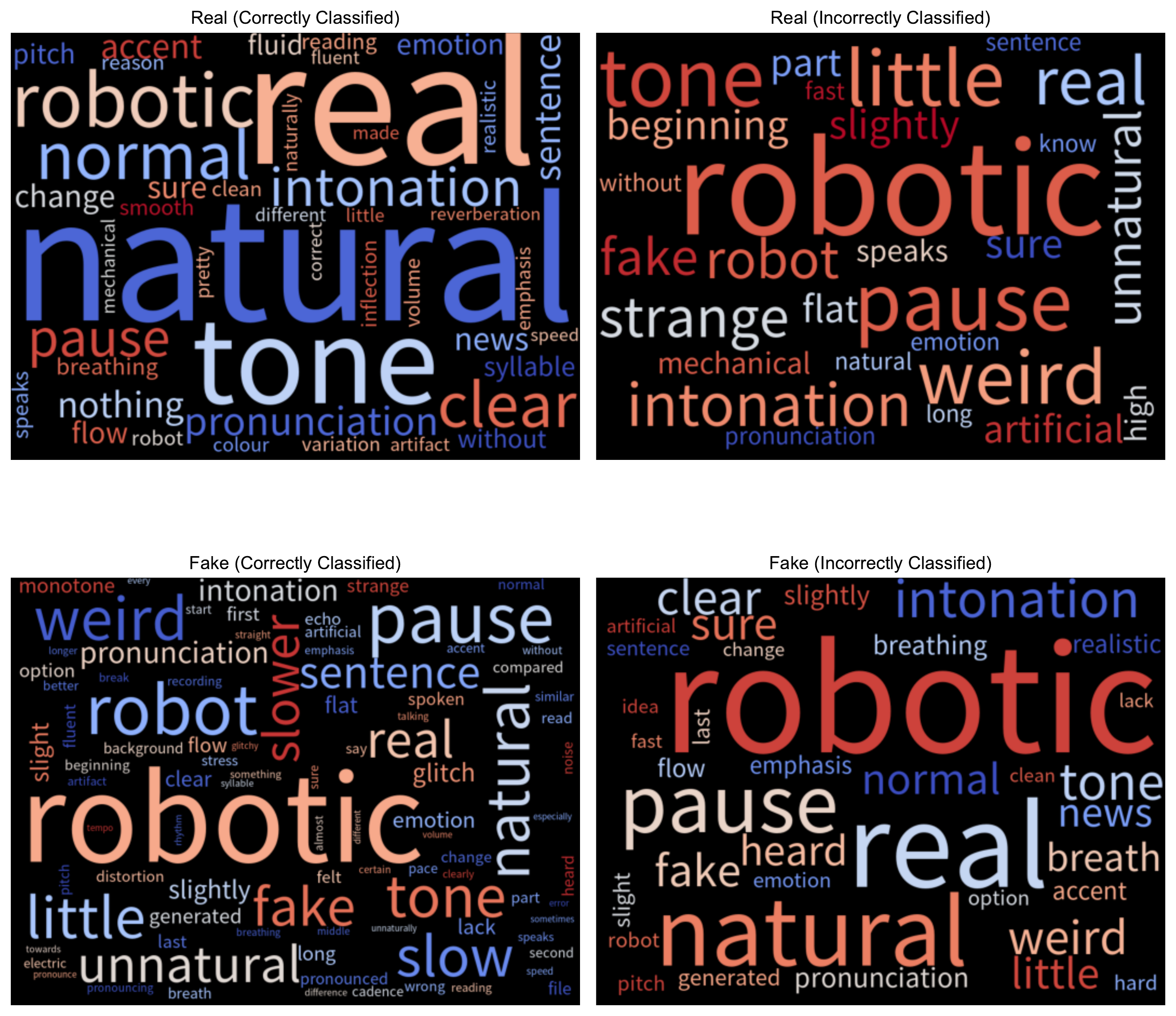}
\caption{{\bf Word clouds containing justifications for the English-language clips.}}
\label{fig9}
\end{figure}

\begin{figure}[!h]
\includegraphics[width=\linewidth]{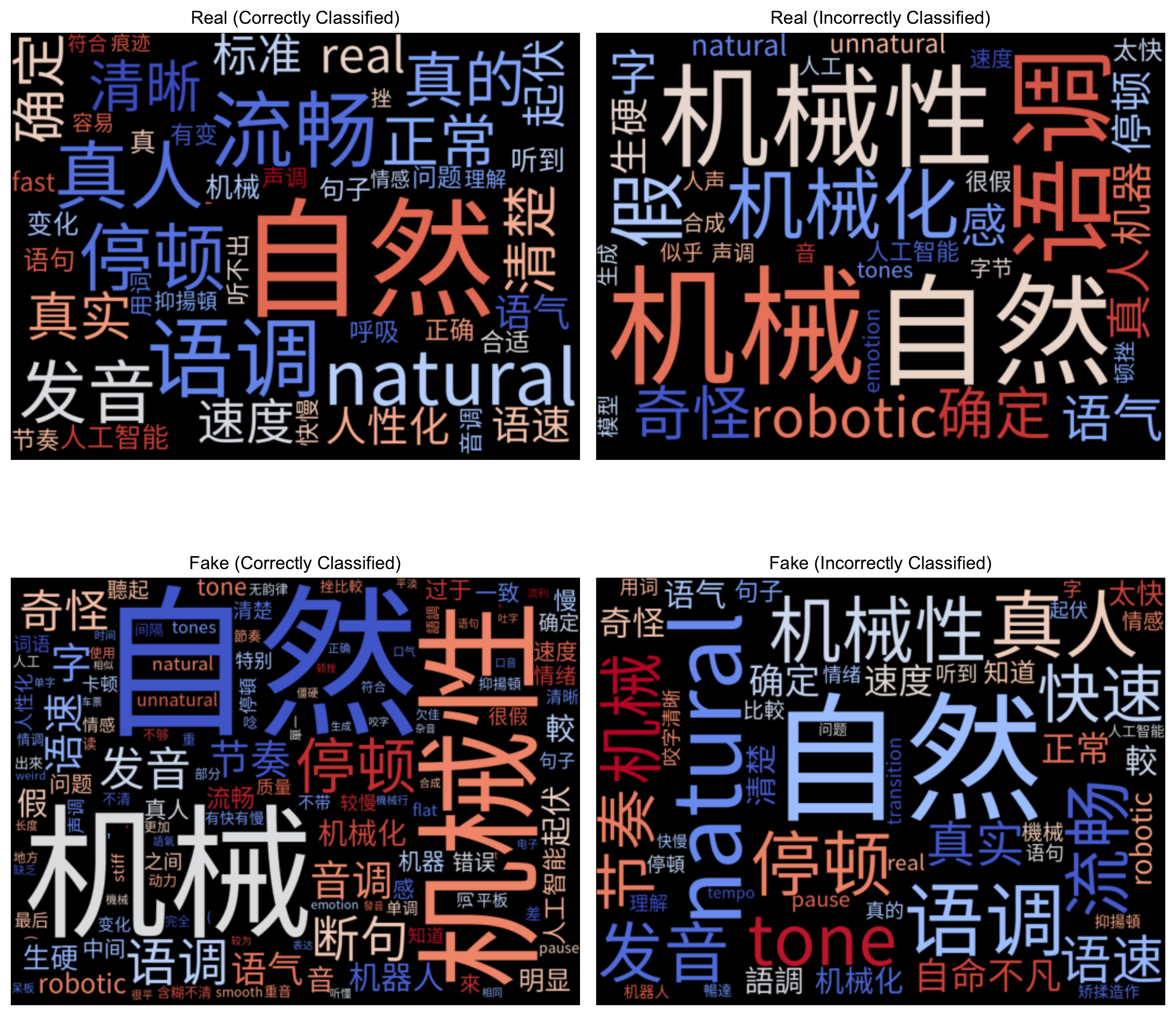}
\caption{{\bf Word clouds containing justifications for the Mandarin-language clips.} Note participants for the Mandarin tasks provided justifications in both Mandarin and English.}
\label{fig10}
\end{figure}

To understand how participants assessed the genuineness of audio clips, we analyzed their freeform text responses. We grouped responses by language, clip authenticity, and whether participants made the correct choice. We then created word clouds using tf-idf weightings. Tf-idf measures the importance of a word within a document compared to a collection of documents to account for frequently appearing words \cite{leskovec2014}. Figs \ref{fig9} and \ref{fig10} show the English and Mandarin word clouds.

\newpage

Participants referred to the same characteristics regardless of whether they made the correct decisions. For example, in Figure \ref{fig9}, participants who correctly classified bona fide utterances as legitimate (in the top left of Figure \ref{fig9}) mentioned pauses, tone and intonation. However, participants who incorrectly categorized bona fide utterances as fake (top right of Figure \ref{fig9}) also referred to these exact attributes. We compared responses by the actual label of the clips and whether participants made the correct response. We did not find substantial differences between these segments. Therefore, automated detectors that incorporate these human characteristics would produce limited improvements. We observed this activity in both English and Mandarin. They tended to rely on intuition to make classifications, referring to naturalness (\chinese{自然}) and robotic (\chinese{机械}) sounds. Beyond intuition, English and Mandarin participants also commonly referenced pauses (\chinese{停顿}), intonation (\chinese{语调}), pronunciation (\chinese{发音}), and speed (\chinese{速度}). 

Regarding differences between languages, there were more references to breathing among the English-speaking participants. In contrast, Mandarin-speaking participants mentioned the speaker's cadence (\chinese{节奏}), pacing between words (\chinese{断句}), and fluency (\chinese{流畅}). This result may be due to differences in timing properties between the two languages. English is stress-timed, while Mandarin is syllable-timed \cite{mok08_speechprosody}.

%% file: 5_limitations.tex
\section*{Limitations}

Although our setup enabled comparison with automated detectors, it does not necessarily reflect more realistic scenarios where a listener may encounter speech deepfakes.

Firstly, the balance of deepfakes we presented in our experiment does not reflect the proportion that occurs in the wild. Participants were equally likely to encounter deepfakes as bona fides in the task. However, AI-generated content (including the use of deepfakes for nefarious purposes) is still rare for now. In addition, we could expect participants to be much more attentive to the occurrence of deepfakes as we informed them about the nature of the task.

Moreover, we minimized contextual information in our stimuli. For example, we do not examine situations where the listeners' contextual knowledge (such as awareness of the speaker's identity, emotional status, the number of parties in a conversation, or political affiliations) may have informed their decisions. These aspects may be relevant to typical use cases where speech deepfakes may arise, such as false news propagation \cite{caldwell2020ai}. Future work could look at exploring how these characteristics influence detection.

Additionally, we asked participants in both languages to listen to utterances purporting to originate from a single female speaker. Given that age and gender of speakers influence speech perception \cite{hummert1998communication, strand1999uncovering}, future work could consider how varying speaker identity affects deepfake detection performance.

To generate our deepfake stimuli, we used an older approach which is not necessarily illustrative of the state-of-the-art speech synthesis algorithms. Although our results indicate how well humans can detect speech deepfakes generated with limited-computational resources, they may not faithfully reflect performance under the most current conditions.

%% file: 6_discussion.tex
\section*{Discussion}

Humans can detect speech deepfakes, but not consistently. They tend to rely on naturalness to identify deepfakes regardless of language. As speech synthesis algorithms improve and become more natural, it will become more difficult for humans to catch speech deepfakes.

Although there are some differences in the features that English and Mandarin speakers use to detect deepfakes, the two groups share many similarities. Therefore, the threat potential of speech deepfakes is consistent despite the language involved.

It will be easier for adversaries to generate more deepfakes as the computational barrier for synthesizing data lowers. More deepfakes in the wild will have a knock-on effect. Adversaries will have more opportunities to scale their operations, particularly for disinformation such as impersonations and spear phishing \cite{mirsky2022threat}.

Ultimately, the battle between deepfake creation and detection is an arms race \cite{chesney2019deepfakes}. How can we defend against falling prey to deepfake trickery? Our binary scenario shows that comparing against reference audio is helpful if we know the speaker's identity. However, we do not always have this information.

Increasing awareness by showing people examples of deepfake audio has a limited effect, as demonstrated by our familiarization results. Spending more time evaluating the clips does not seem to help either.

To summarize, attempting to improve human detection capabilities is unrealistic. We show that even in a controlled environment where the task is easier (participants are aware of the presence of speech deepfakes and the deepfakes are not created using state-of-the-art speech synthesizers), deepfake detection is not high. Our results suggest the need for automated detectors to mitigate a human listener's weaknesses. Automated detectors' performance on in-domain data indicates they can pick up on subtleties that humans cannot. However, we show they are brittle and fail to work when there are changes in the test audio's environmental conditions. Given the extent of human limitations and the increasing availability of computational resources for deploying detectors, research should focus on improving these detectors. In the meantime, crowd-sourcing is a reasonable mitigation. We confirm crowd performance is on par with the top-performing automated detectors and is not as brittle. Extending fact-checking tools to include audio evaluations is one way to protect against deepfake threats.

%% file: 7_supporting.tex
\section*{Supporting information}
\label{supporting}

\renewcommand{\thefigure}{S\arabic{figure}}
\setcounter{figure}{0}

\begin{figure}[!h]
\includegraphics[width=\textwidth]{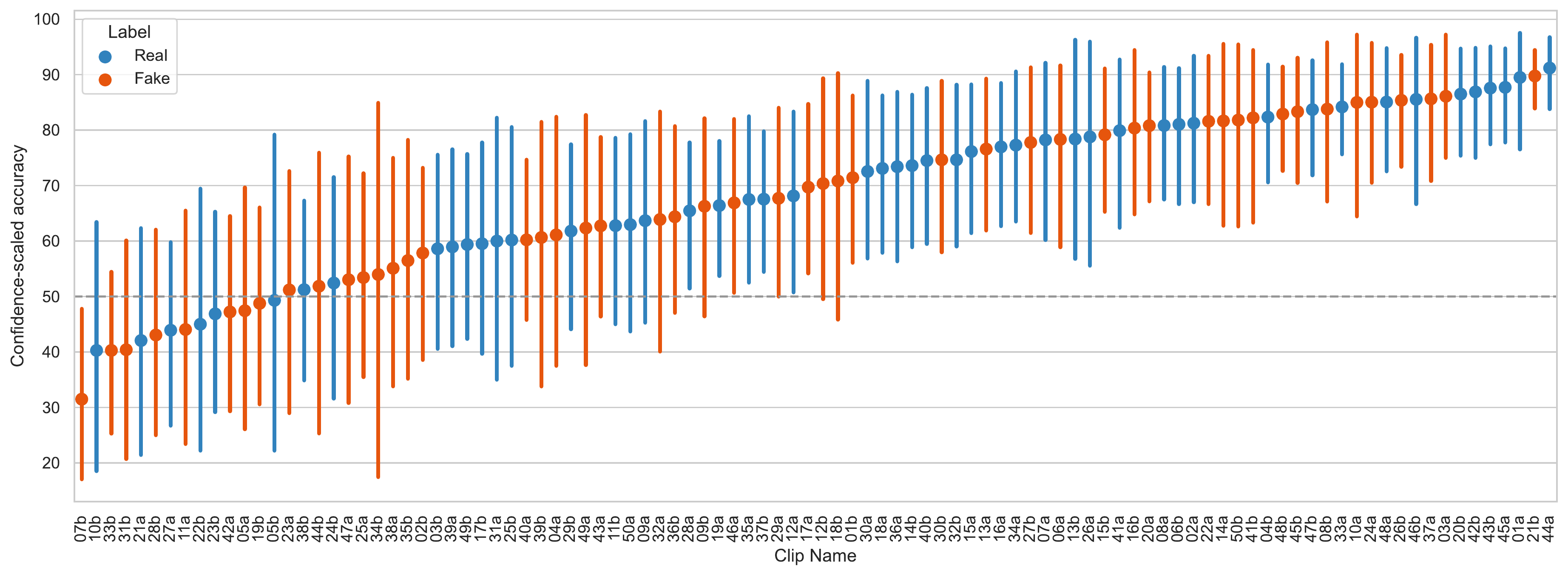}
\caption{Confidence-adjusted accuracy scores per clip (English, unary, no familiarization).}
    \label{S1_Fig}
\end{figure}

\begin{figure}[!h]
\includegraphics[width=\textwidth]{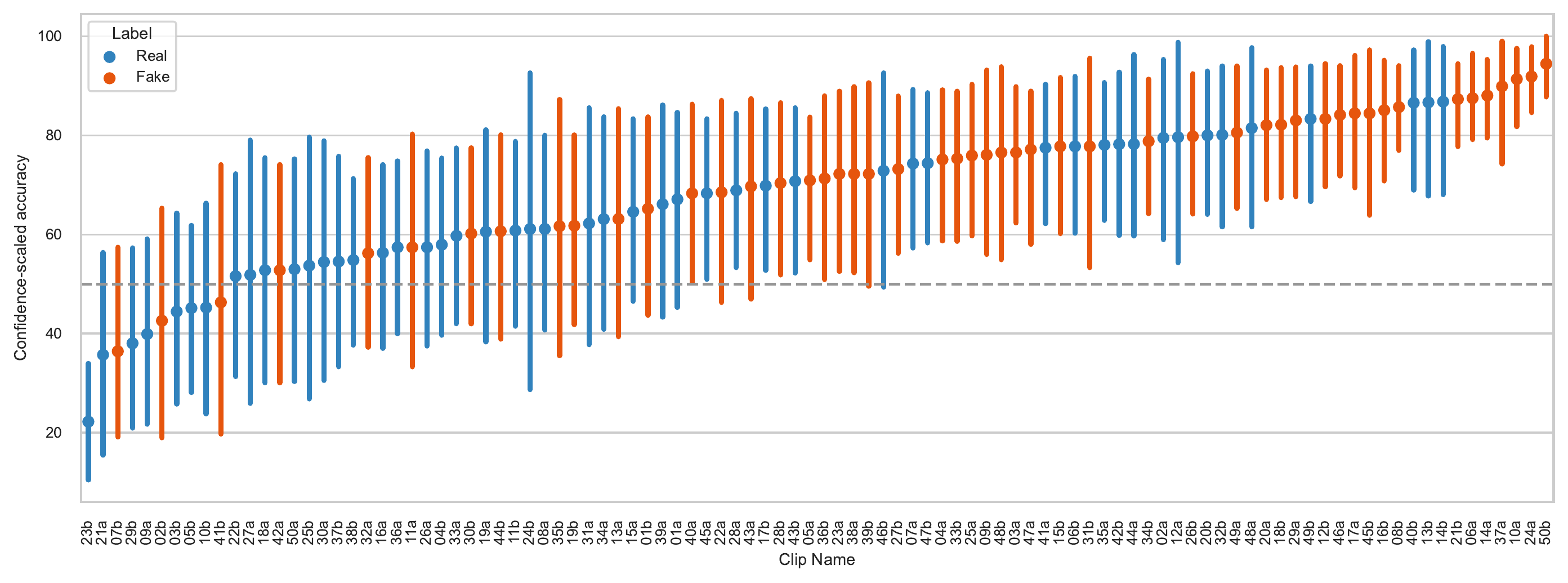}
\caption{Confidence-adjusted accuracy scores per clip (English, unary, familiarization).}
    \label{S2_Fig}
\end{figure}

\begin{figure}[!h]
\includegraphics[width=\textwidth]{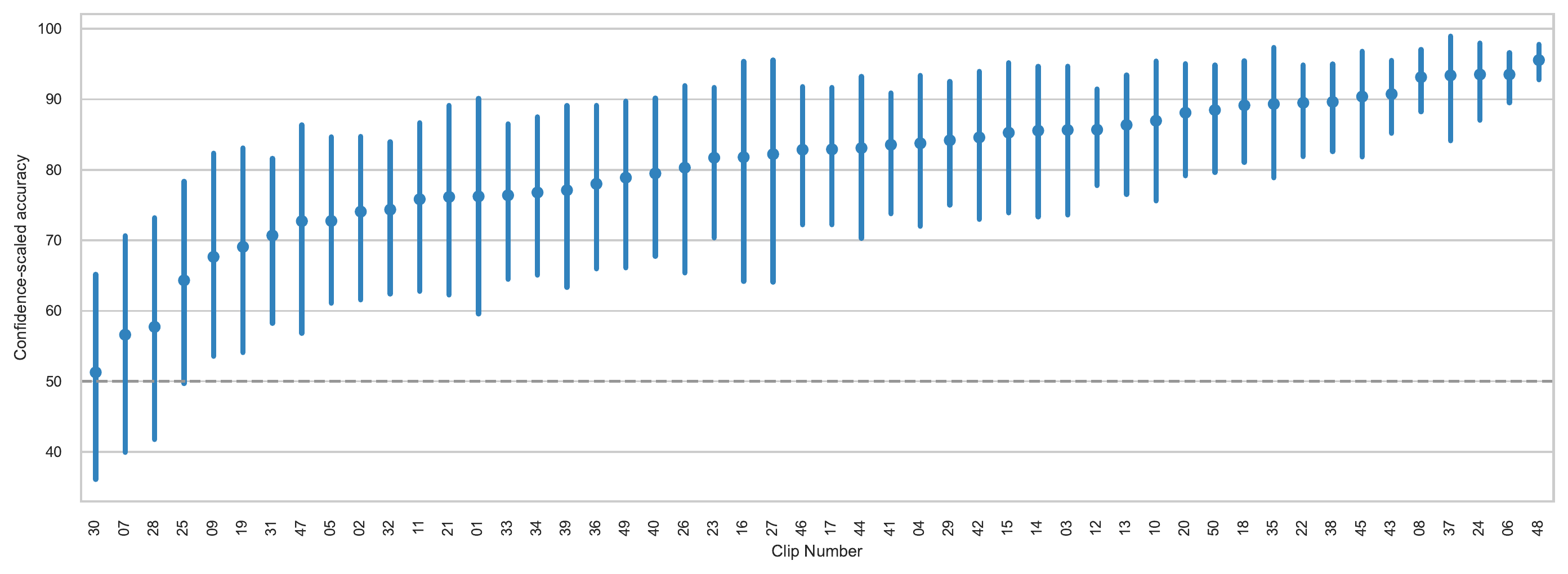}
\caption{Confidence-adjusted accuracy scores per clip (English, binary, no familiarization).}
    \label{S3_Fig}
\end{figure}

 \begin{figure}[!h]
\includegraphics[width=\textwidth]{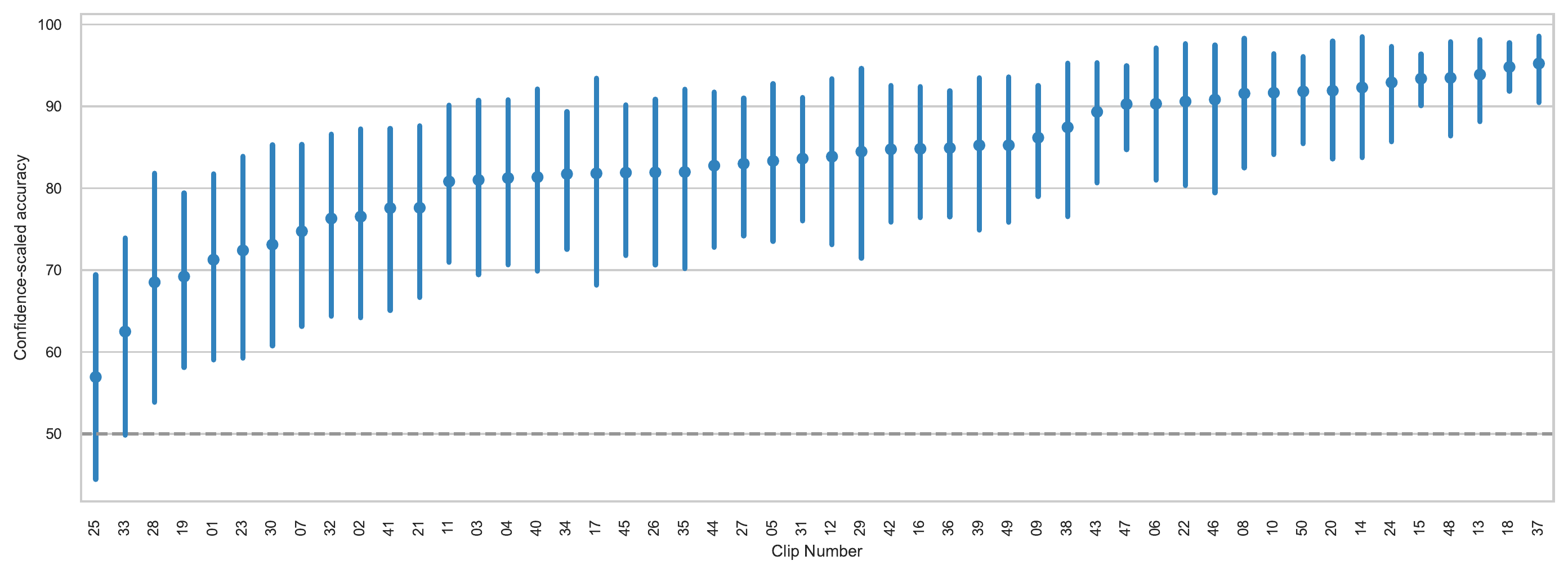}
\caption{Confidence-adjusted accuracy scores per clip (English, binary, familiarization).}
    \label{S4_Fig}
\end{figure}
 
 \begin{figure}[!h]
\includegraphics[width=\textwidth]{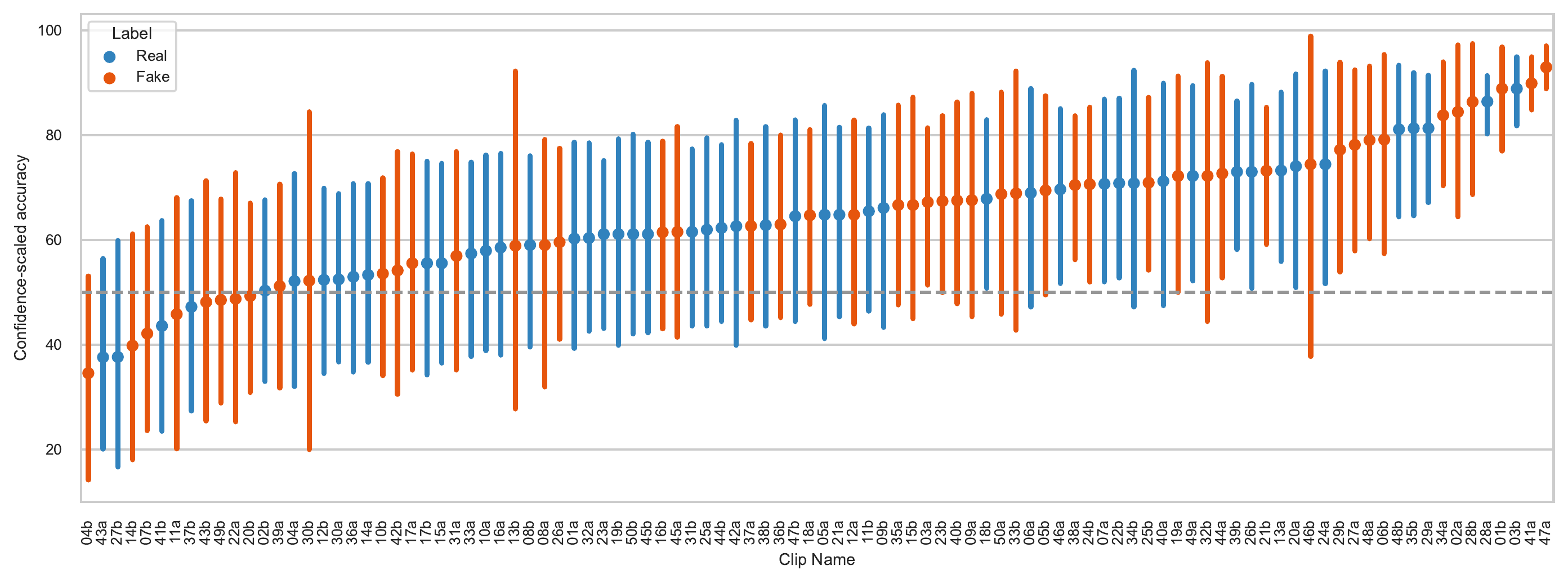}
\caption{Confidence-adjusted accuracy scores per clip (Mandarin, unary, no familiarization).}
    \label{S5_Fig}
\end{figure}

 \begin{figure}[!h]
\includegraphics[width=\textwidth]{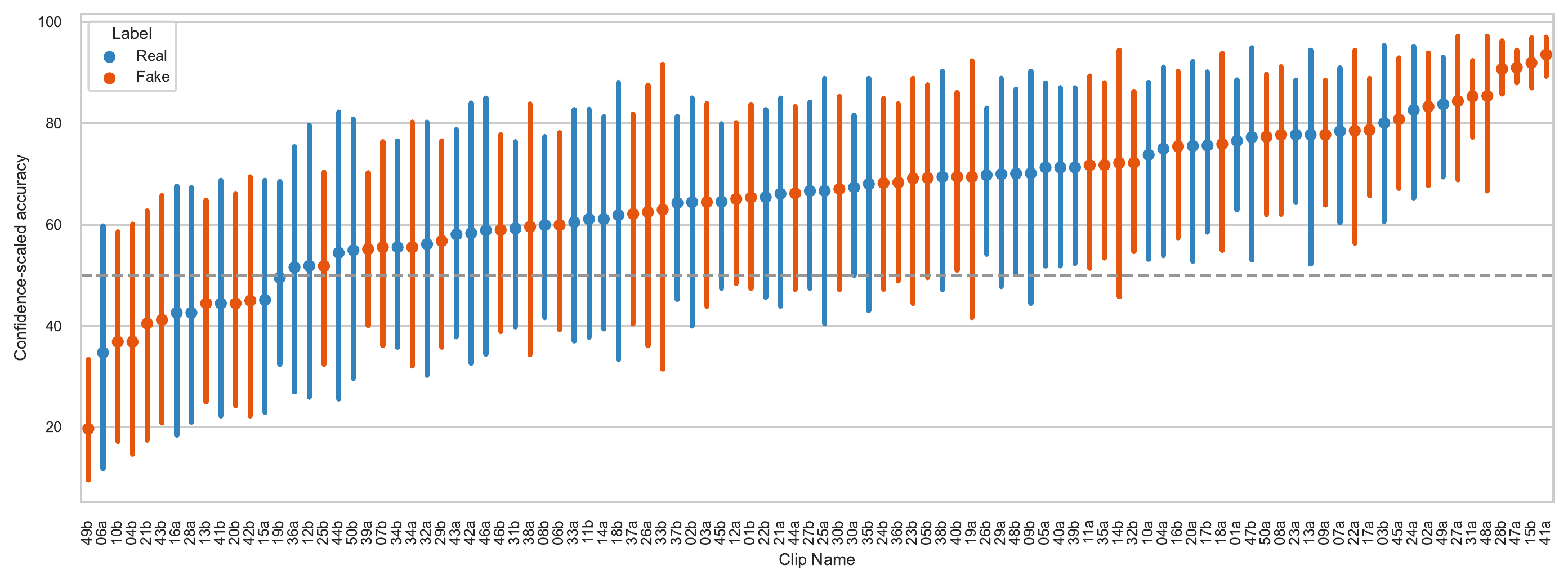}
\caption{Confidence-adjusted accuracy scores per clip (Mandarin, unary, familiarization).}
    \label{S6_Fig}
\end{figure}

 \begin{figure}[!h]
\includegraphics[width=\textwidth]{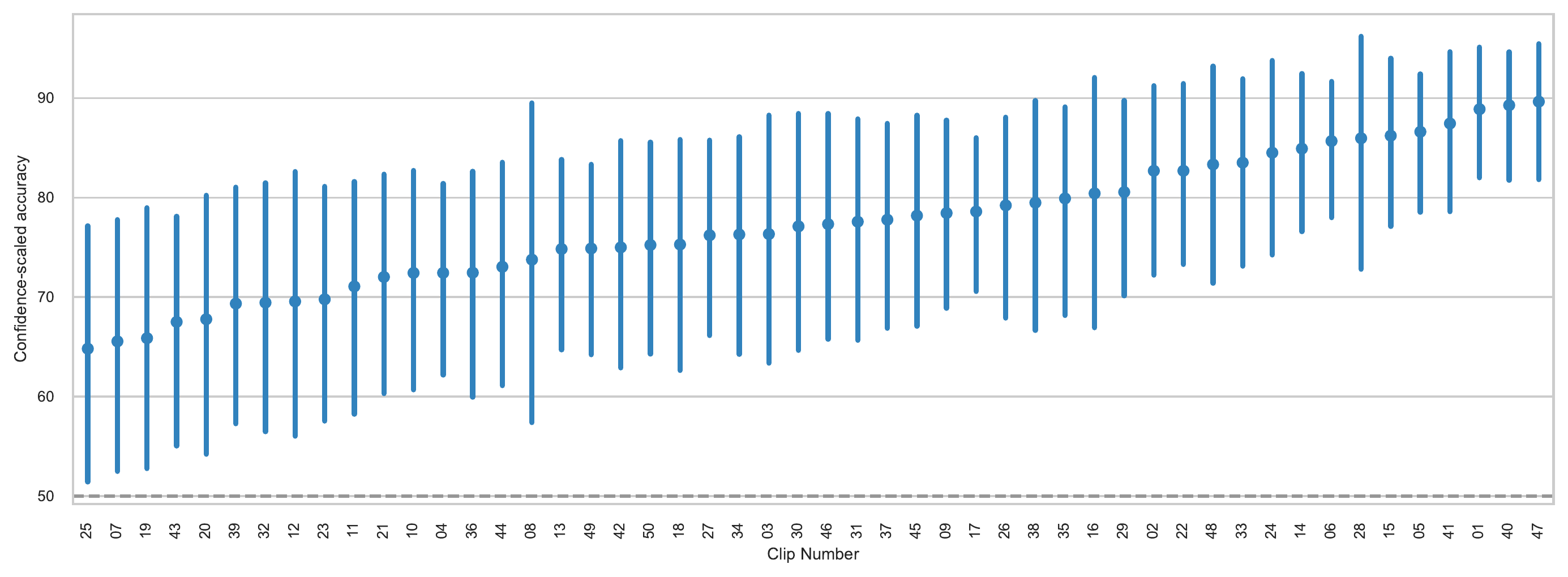}
\caption{Confidence-adjusted accuracy scores per clip (Mandarin, binary, no familiarization).}
    \label{S7_Fig}
\end{figure}

 \begin{figure}[!h]
\includegraphics[width=\textwidth]{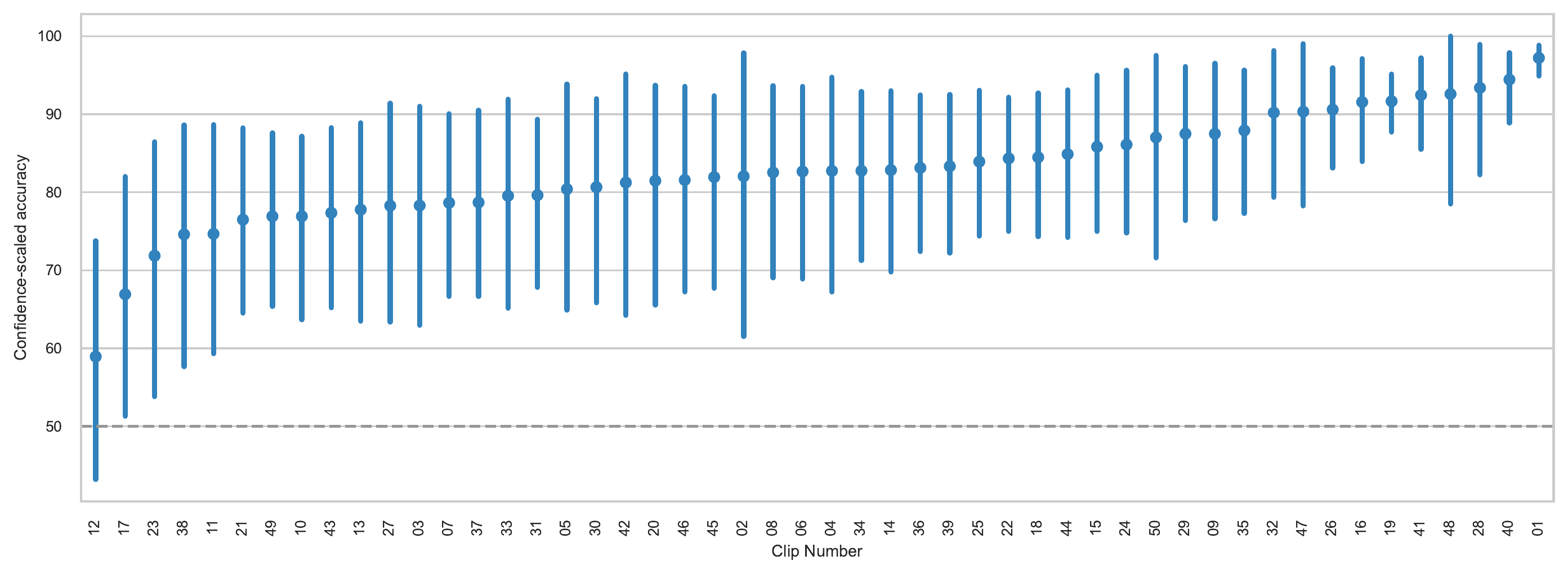}
\caption{Confidence-adjusted accuracy scores per clip (Mandarin, binary, familiarization).}
    \label{S8_Fig}
\end{figure}
\clearpage